# Daytime sub-ambient radiative cooling with vivid structural colors mediated by coupled nanocavities

Shenghao Jin,[1] Ming Xiao,[2] Wenbin Zhang,[1] Boxiang Wang,[1, *] and Changying Zhao[1, *]

1 Institute of Engineering Thermophysics, School of Mechanical Engineering, Shanghai Jiao Tong University, Shanghai 200240, China
2 College of Polymer Science and Engineering, Sichuan University, Chengdu, 610065, China
Correspondence: **changying.zhao@sjtu.edu.cn** (Changying Zhao) and **wangboxiang@sjtu.edu.cn** (Boxiang Wang)

## Abstract

Daytime radiative cooling is a promising passive cooling technology for combating global warming. Existing daytime radiative coolers usually show whitish colors due to their broadband high solar reflectivity, which severely impedes applications in real-life situations with aesthetic demands and effective display. However, there is a trade-off between vivid colors and high cooling performance because colors are often produced by absorption of visible light, decreasing net cooling power. To break this trade-off, we design multilayered structures with coupled nanocavities and produce structural colors with high cooling performance. Using this design, we can obtain colorful radiative coolers which show a larger color gamut (occupying 17.7% sRGB area) than reported ones. We further fabricate colorful multilayered radiative coolers (CMRCs) and demonstrate they have temperature drops of 3.4 ℃~4.4 ℃ on average based on outdoor experiments. These CMRCs are promising in thermal management of electronic/optoelectronic devices and outdoor facilities.

Keywords: structural color, sub-ambient daytime radiative cooling, coupled nanocavities, multilayered structure

# Introduction

Nowadays, the demand for cooling is significantly increasing due to the climate change caused by global warming.[1] Among many others, daytime radiative cooling is a promising passive cooling technology that can realize sub-ambient cooling of terrestrial objects. To achieve sub-ambient temperature under direct sunlight, a daytime radiative cooler requires a high solar reflectivity (0.3-4 μm) and high emissivity at mid-infrared (MIR) wavelengths, especially within the primary atmospheric window (8−13 μm).[2] Since Raman et al.[3] experimentally realized sub-ambient daytime radiative cooling under direct sunlight with a multilayered film in 2014, various daytime radiative coolers have been demonstrated in recent years, including metal-based photonic devices,[3]-[5] porous polymers,[6]-[8] and polymer-dielectric composites.[9]-[16] These reported radiative coolers show white or silver colors due to their broadband high reflectivity in the solar spectrum. This whitish color is not desirable for aesthetic and functional reasons,[17]-[19] which severely restricts their uses in wearable devices,[20] functional clothes for personal thermal management,[21] eco-friendly buildings,[16] and other outdoor facilities.[22]

It is challenging to design radiative coolers with vivid colors because coloring a radiative cooler decreases its radiative cooling performance due to the significant increase in the thermal load when spectral absorption for coloration is introduced in the visible spectrum.[23] Recently, a variety of works attempted to design colorful radiative coolers by adding chemical absorbers (pigments, dyes, etc.) or photoluminescent materials (like quantum dots).[24]-[25] However, these approaches result in a large reduction of net cooling power because the broad and uncontrollable absorption peaks cause excessive solar power absorption. For example, the pigmentary coloration of porous P(VdF-HFP)–based radiative cooling bilayers resulted in 5~20°C increase compared to the ambient temperature,[24] which largely reduced the sub-ambient cooling ability. Inappropriate coloration will cause a radiative cooler to lose its function of sub-ambient daytime cooling. To design colorful radiative coolers, we need to break the trade-off between vivid colors and high net cooling powers.

To this end, we use structural colors. Structural colors are generated from the interference, diffraction, or scattering caused by the interaction of light with micro/nano structures,[26]-[28] which is fundamentally different from pigmentary and photoluminescent colors resulting from the energy level transition of electrons at specific wavelengths. By delicately manipulating micro/nano-structures, one can achieve highly controllable structural colors with large-gamut, high-saturation, high-brightness, and high-resolution.[29]-[32] Moreover, compared to pigmentary and photoluminescent colors, structural colors are more easily tuned with controlled narrowband optical resonances without introduction of excessive solar power absorption. Another advantage of structural colors is that they are resistant to photo-bleaching and chemical washing.[33] Therefore, radiative cooling with structural color is very suitable for addressing the restrictions of colorful radiative coolers for practical application.

Several works have demonstrated radiative cooling with structural color by optical resonant modes such as localized surface plasmon resonance and Bragg diffraction.[34]-[38] However, the absorption or transmission of solar radiation due to these resonances is still large, resulting in low solar reflectivity and forbidding the realization of sub-ambient daytime radiative cooling. Therefore, a judicious photonic design of micro- or nanostructures is critical to minimize solar absorption for high net cooling power while achieving vivid structural color. Thin films stacked by 1D metal−insulator−metal (MIM) are promising candidates for structurally colorful radiative cooling. This structure usually exhibits a single narrow reflection dip caused by a narrow absorption peak due to the optical resonant mode inside.[39]-[41] Lee et al.[42] and Xi et al.[43] have reported colorful radiative coolers with 1D MIM structures. However, the colors of these radiative coolers are not saturated enough due to only one narrow reflection dip of MIM structure within the visible region. Blandre et al. used 1D MIMIM (metal−insulator−metal-insulator-metal) to make more saturated colorful radiative coolers,[44] but there is still a large amount of solar power absorption about ~170 W /

m² due to the broadband absorption peak, which greatly limits the net cooling power (around ~5 W / m² under the theoretical limit of daytime radiative cooling ability [45]). Under this circumstance, the sub-ambient cooling can hardly be realized. So far, there is still no photonic structure that can successfully break the trade-off between saturated colors and daytime radiative cooling. Meanwhile, the existing sub-ambient radiative cooler has a very limited color gamut, which indicates an insufficient coloration capability.

Herein, we design a series of integrated colorful multilayered radiative coolers (CMRCs) where a selective emitter is stacked on the top of a photonic structure. The emitter is made of $SiO_2$ and $Si_3N_4$ layers, and the photonic structure for coloration is made of two nanocavities consisting of $SiO_2$, $TiO_2$, and silver layers. By exploiting the coupling effect between two optical resonances inside the stacked nanocavities, we can create one or two narrow reflection dips, via modulating the coupling strength, to produce a full spectrum of colors while maintaining high solar reflectivity ($R_{ave}$ = 85.9% − 95.9%). With optical simulations and thermal analyses, we demonstrate that the CMRCs can present a much larger color gamut (17.7% of sRGB) than all previous works with sub-ambient daytime radiative cooling. We also experimentally fabricated CMRCs which have higher sub-ambient cooling performance than other reported colorful radiative coolers. These advantages give CMRCs the potential to be utilized in aesthetic decorations and passive thermal management for smartphones, wearable electronic/optoelectronic devices, and outdoor facilities.

## Principle of the CMRC design

To achieve vivid colors and sub-ambient daytime cooling, we demonstrate a colorful multilayered radiative cooler consisting of a selective emitter on the top and a photonic structure on the bottom (Figure 1a). The selective emitter is made of one $SiO_2$ layer and one $Si_3N_4$ layer, which have superior solar transparency, and intrinsically high emissivity in the atmospheric window (8~13 μm).[46]-[48] The emitter can realize broadband high emissivity in the atmospheric window for cooling with total optimized thicknesses from 1.25 μm to 1.32 μm (The optimization method is shown in Supporting Information Section 1 and Table S1). The photonic structure comprises a stacked multilayer with a quasi-distributed Bragg reflector (qDBR) on the top and a MIM nanocavity on the bottom. The qDBR consists of two periods of $SiO_2/TiO_2$ structure having dielectric contrast of ~2/3 ($n_{SiO_2}/n_{TiO_2}$). We dub it a qDBR because this structure does not have a complete bandgap like a real DBR (1D photonic crystal) since the number of periods is small. MIM nanocavity consists of silver as the metal material to reflect solar energy and $SiO_2$ as the lossless insulator material.

It is found that qDBR-Ag structure and MIM can exhibit optical resonant modes inside dielectric cavities by themselves (Figure 1b and Figure S1a-b). When stacking these two structures together, two coupled resonators will contribute to the combined optical response, which can be utilized to realize structural coloration with high average solar reflectivity. For instance, when MIM and qDBR-Ag are stacked via one thin central silver layer (Figure 1b), narrow-width resonance (MIM) and broad-width resonance (qDBR-Ag) combines to asymmetric narrow double-reflection dips, whose lineshapes are determined by a coupling effect. Thanks to the coupling effect, the intrinsic material loss of metallic layers in qDBR-MIM structure can be largely avoided since energy is efficiently exchanged between the two nanocavities without being significantly absorbed. Therefore, the qDBR-MIM nanocavity has higher reflectivity at the resonant frequencies compared to two composing photonic structures (Figure 1b), which is beneficial to maintain high average solar reflection when displaying vivid structural colors. Under the mode coupling, the qDBR-MIM nanocavity has one more narrow reflection dip for coloration, producing higher color saturation compared to MIM. It also has narrower reflection dips than the original qDBR-Ag structure, so the qDBR-MIM nanocavity has a higher average solar reflectivity (Figure 1b). Thus, the coupled nanocavities integrate the advantages of two separate structures and overcome their shortcomings. Moreover, the coupling effect can be flexibly tuned by manipulating the geometric parameters, which can adjust the spectral positions and lineshapes of two resonances without affecting the reflection at other wavelengths. These abilities allow our structures to have significantly higher average solar reflection compared with the single-resonance-based photonic structures when displaying specific colors.

To further demonstrate the advantages of our qDBR-MIM nanocavity in coloring radiative cooler, we design three colorful radiative coolers (CRCs) with identical selective emitters for cooling and different photonic structures (MIM, qDBR-metal, and qDBR-MIM) for coloration (Figure 1c), which can display similar pink colors (Figure 1d). The CRC3 with qDBR-MIM has the highest average solar reflectivity since the coupled nanocavities can display high-saturation structural colors by tuning the coupling strength of the two optical resonances without impairing the reflectivity in other wavelengths. On the contrary, to achieve a similar color, one needs to reduce the thickness of the metal reflective layer in the MIM and qDBR-metal structures, which would broaden the linewidth of the single resonance peak and inevitably reduce the average solar reflectivity (Table S2). Therefore, by manipulating the geometric parameters (thickness of each layer) to tune the coupling strength, we can obtain a rich gamut of structural colors from double or single narrow reflection dips, which will avoid excessive solar power absorption and ensure cooling performance.

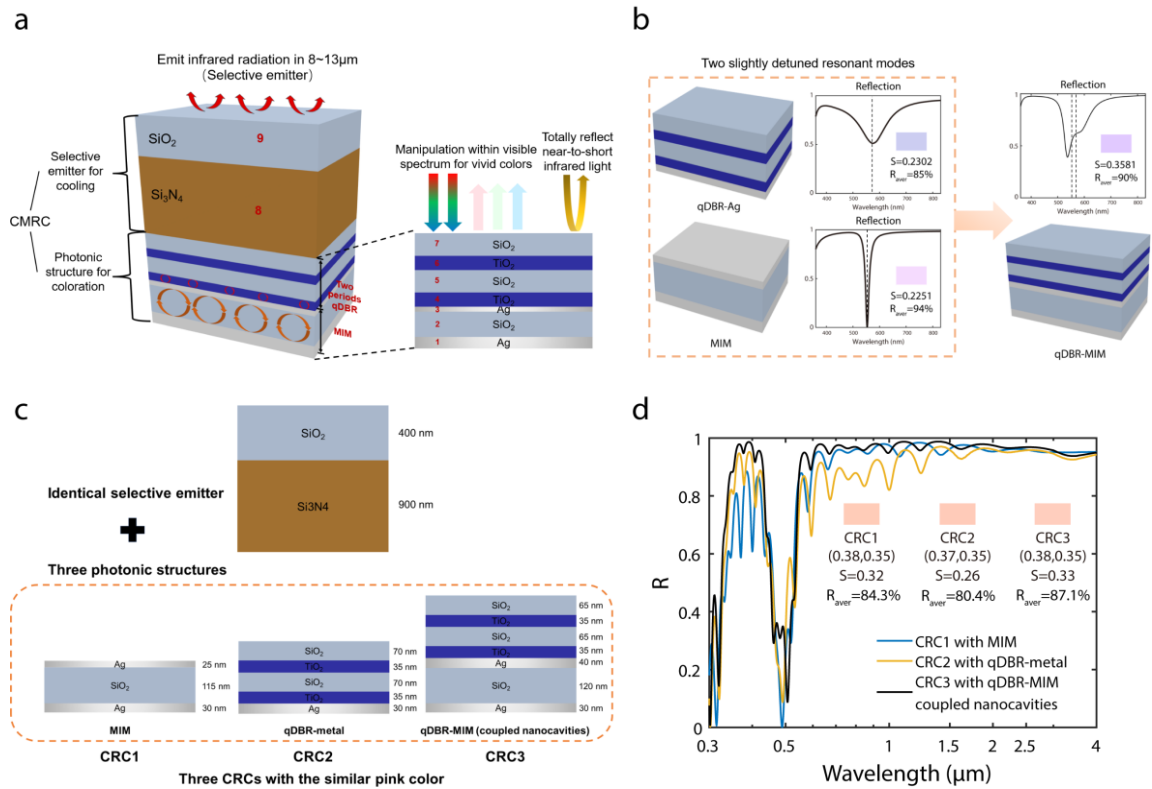

**Figure 1**. Design principle of the colorful daytime radiative cooler. (a) Schematic

illustration of a colorful multilayered radiative cooler showing chromatic color with high reflectivity, which consists of a selective emitter and a pair of coupled nanocavities (qDBR-MIM nanocavity) as photonic structure. To facilitate the subsequent analysis, we numbered the multilayered structure from bottom to top. (b) Comparison of color saturation and average solar reflectivity between qDBR-MIM nanocavity and original structures (MIM and qDBR-Ag structure): MIM nanocavity has a higher average solar reflectivity, but less color saturation, while qDBR-Ag structure has a higher color saturation, but lower average solar reflectivity. After connecting by the central Ag layer (Layer 3), qDBR-MIM nanocavity will exhibit higher color saturation and considerable average solar reflectivity under the coupling of two resonant modes compared to two original structures. (c) Schematic illustration of three colorful radiative coolers (CRCs) with identical selective emitters and different photonic structures for coloration. Three CRCs display the similar pink color. (d) Comparison of average solar reflectivity between CRCs with different photonic structures. CRC3 with qDBR-MIM coupled nanocavities has the highest average solar reflectivity when displaying the similar color.

The thickness of the central metal layer (Layer 3 in Figure 1a) determines the coupling strength of two resonant modes when the stacked qDBR-MIM structure has fixed top qDBR layers and bottom dielectric cavity thicknesses.[49]-[50] We can tune thickness Layer 3 to obtain different lineshapes of reflective spectra. For instance, when the individual resonant frequencies of two independent resonators are similar or only slightly detuned to each other, two absorption peaks can be observed as long as the Layer 3 is thin (Figure 2a). The resonant frequencies of stacked qDBR-MIM structure are completely different from the resonant frequencies of two independent structures, due to the mode splitting. The details of the mode coupling analysis of this structure by coupled harmonic-oscillator model are provided in Supporting Information Section 3. After comparing the Rabi-splitting frequency $\Omega$ and the linewidths of two original composing structures $(\gamma_{\text{qDBR}-\text{Ag}} + \gamma_{\text{MIM}})/2$, we can find that two optical resonances inside the qDBR-MIM structure (shown in Figure 2a) exhibit a strong coupling behavior (Figure S2). Meanwhile, as shown in Figure 2a and Figure S3, the out-of-phase (odd) and in-phase (even) superposition of the resonant modes corresponding to the antisymmetric and symmetric modes in the strong coupling regime can be observed clearly. In addition, the intrinsic material loss in the Layer 3 and Layer 1 of the qDBR-MIM structure determining the optical power absorption at resonant frequencies can be greatly suppressed since the energy is efficiently exchanged between the two nanocavities without being absorbed (Figure S3). This phenomenon can significantly reduce solar absorption while displaying vivid colors. Moreover, with the increase of Layer 3 thickness from 10 nm to 80 nm, the coupling strength would gradually decrease and the two resonant modes would eventually become decoupled with each other, resulting in a single-absorption-peak spectrum (Figure 2c). This phenomenon will directly change the color display of the structure from green to magenta (Figure 2b). We also quantitatively calculate the influence of the Layer 3 thickness on the coupling strength. With the increase of Layer 3 thickness from 20 nm to 50 nm, the Rabi splitting energy decreases from 569.2 meV to 0 meV, which means the coupling between two resonant modes changes from strong coupling to weak coupling regime, and eventually they become decoupled (Figure S4).

For a stacked qDBR-MIM structure with varied thickness of Layer 2, the lineshape of reflectivity spectrum also changes since the resonant frequency of nanocavity is mainly determined by the thickness of its dielectric layer. When the resonant frequencies of two independent resonators are highly detuned, we observe completely different absorption spectrum appearance and electric field distribution compared to those under strong coupling (Figure 2d). Two resonant peaks of decoupled nanocavity are almost identical to the independent resonators (1st peak: 494 nm, original qDBR-Ag: 503 nm. 2nd peak: 647 nm, original MIM: 641 nm.). Meanwhile, the electric field distribution at resonance frequencies of decoupled nanocavity is identical to the independent resonators (Figure S5), which indicates that the hybridization of the optical mode hardly occurs. (Figure 2d). With the increase of Layer 2 thickness from 80 nm to 170 nm, the colors of stacked structures firstly turn from pink to magenta and then turn to cyan (Figure 2e). Meanwhile, a mode anti-crossing behavior can be observed in of qDBR-MIM structures with different thicknesses of Layer 2 (80 nm to 170 nm at intervals of 10 nm, from bottom to the top). The color of each absorption curve is the perceived color of corresponding structure calculated by their spectral reflectivity based on CIE 1931 standard colorimetric system.[51] (f) Mode anti-crossing in a qDBR-MIM structures with fixed thicknesses of qDBR layers and metal layers, whereas the thickness of Layer 2 is varied.

Figure 2f. The shift of symmetric resonant mode from unperturbed frequency (two dash lines) is smaller than antisymmetric resonant mode, which also indicates the occurrence of an asymmetric mode splitting, similar to the symmetric MIMIM nanocavity.[49]

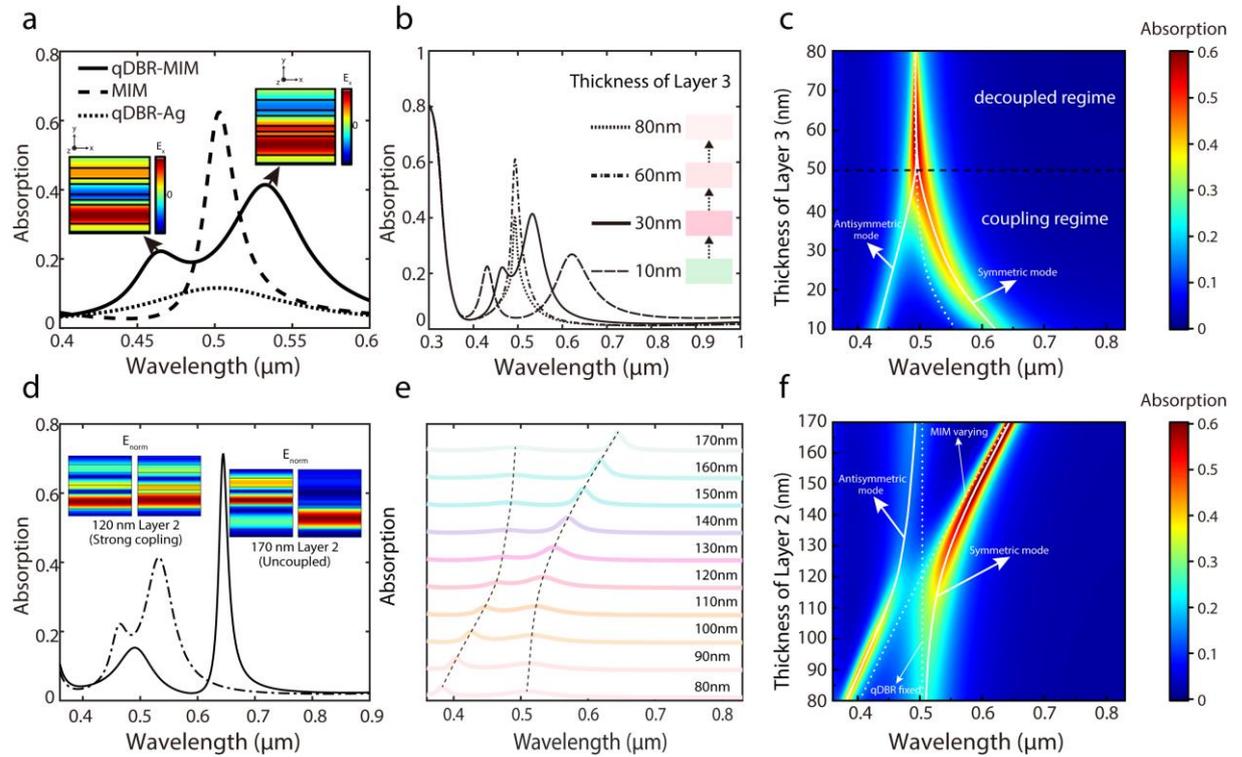

**Figure 2**. Mode coupling in the qDBR-MIM structures. (a) The appearance of absorption spectrum and *x*-component of the electric field at resonant frequencies under the strong coupling. (b) Absorption curves of qDBR-MIM structures with different thicknesses of Layer 3 when the individual resonant frequencies of two independent resonators are similar or only slightly detuned to each other. (c) Resonant wavelengths of qDBR-MIM structures with different thicknesses of Layer 3 when the resonant frequencies of two independent resonators are similar to each other. The mode splitting can be tuned from strong coupling regime to decoupling. (d) The appearance of absorption spectrum with Layer 2 at thickness of 120 nm (dash line) and 170 nm (solid line), and magnitude of electric field at corresponding resonant frequencies (Left: $E_{norm}$ at 1st resonant peak. Right: $E_{norm}$ at 2nd resonant peak). The difference between coupled and decoupled nanocavity can be clearly observed via the electric field distribution at resonant frequencies. Note the thicknesses of one period qDBR are fixed at 65 nm/ 35 nm and the thickness of Layer 3 is fixed at 30 nm. (e) Absorption curves

We can make different structural colors by tuning the geometric parameters of qDBR-MIM, which include the thicknesses of qDBR layers and dielectric layer in MIM nanocavity determining the original resonant frequencies, and the thickness of the central metal (Ag) layer determining the coupling strength of two resonant modes. Compared to independent MIM nanocavities with a narrow single-reflection dip, qDBR-MIM multilayered structure can produce a larger area of color gamut due to more significant spectral contrast in the reflection.

## Simulation

To predict the cooling performance, we establish a model that can evaluate the radiative cooling performance of objects (Supporting Information Section 4). We firstly choose six CMRCs with relatively high color saturation and three CMRCs with relatively high solar reflectivity as examples (Figure S6). These CMRCs have temperature drops of $3°C \sim 8.5°C$ when the net cooling power is zero, and net cooling power of $31 - 85$ W / m² at $T_s = T_{amb} = 30$ °C (Figure S6g-i). Thus, we realize considerable net radiative cooling power with chromatic colors using these CMRCs.

To further illustrate the color and cooling performance of CMRCs, we calculate the net cooling power and corresponding colors (Details are provided in Supporting Information Section 5) of 384 different CMRCs (Figure 3a). To ensure enough solar energy reflection for sub-ambient radiative cooling, the thickness of Layer 1 is fixed at 50 nm. All of the 384 structures have positive net radiative cooling power from 19  W /

m² to 91 W / m² at $T_s = T_{amb} = 30$ °C. To clearly exhibit the gamut of structural colors, we calculate their chromaticity coordinates and plot them $(x, y)$ in CIE 1931 chromaticity diagram (Figure S9a). We define two colors are different when the CIELAB color difference $\Delta E$ is larger than 1.[52] We can obtain 256 different colors from 384 structures. The six colors shown in Figure S6a and Figure S6b are at the edge of color gamut (white and red dots shown in Figure S9a).

By tuning the thickness of Layer 1 (bottom Ag layer) that determines the overall reflection and color of a CMRC, we can delineate the size of color gamut of our CMRCs with sub-ambient cooling ability (net cooling power $P_{net} > 0$ when $T_s = T_{amb}$). To understand how the thickness of Layer 1 affects the cooling performance and color saturation, we tune the thickness of Layer 1 from 30 nm to 80 nm. The thinner Layer 1 is, the higher the corresponding color saturation is (Figure 3b). The corresponding radiative cooling performance will be impaired due to the increase in solar power absorption, which clearly demonstrates the tradeoff between coloration and daytime radiative cooling. The net cooling power and color saturation are quite considerable at the Layer 1 thickness of 50 nm, while both of them tend to change slowly when the thickness of Layer 1 exceeds 50 nm. Furthermore, we exhibit the color gamut determined by CMRCs with positive net cooling power at ambient among 1920 types of structures (Figure 3c). Compared to previous works,[25],[34],[42],[44] our designed CMRCs present a much larger color gamut with sub-ambient daytime radiative cooling ability. They can occupy 17.7% area of sRGB gamut (Figure 3c), which has not been realized so far for colorful radiative coolers with sub-ambient daytime cooling ability. Moreover, our CMRCs can achieve a much larger color gamut with sub-ambient cooling ability than the original composing structures by tailoring the coupling effect (Figure S9c). We also compare the cooling performance of three CRCs shown in Figure 1c (Figure S10). Due to the highly flexible manipulation of the coupling effect in the qDBR-MIM structure, our designed CMRC has much better cooling performance than the RCs colorized by MIM structure and qDBR-metal structure when displaying similar pink colors.

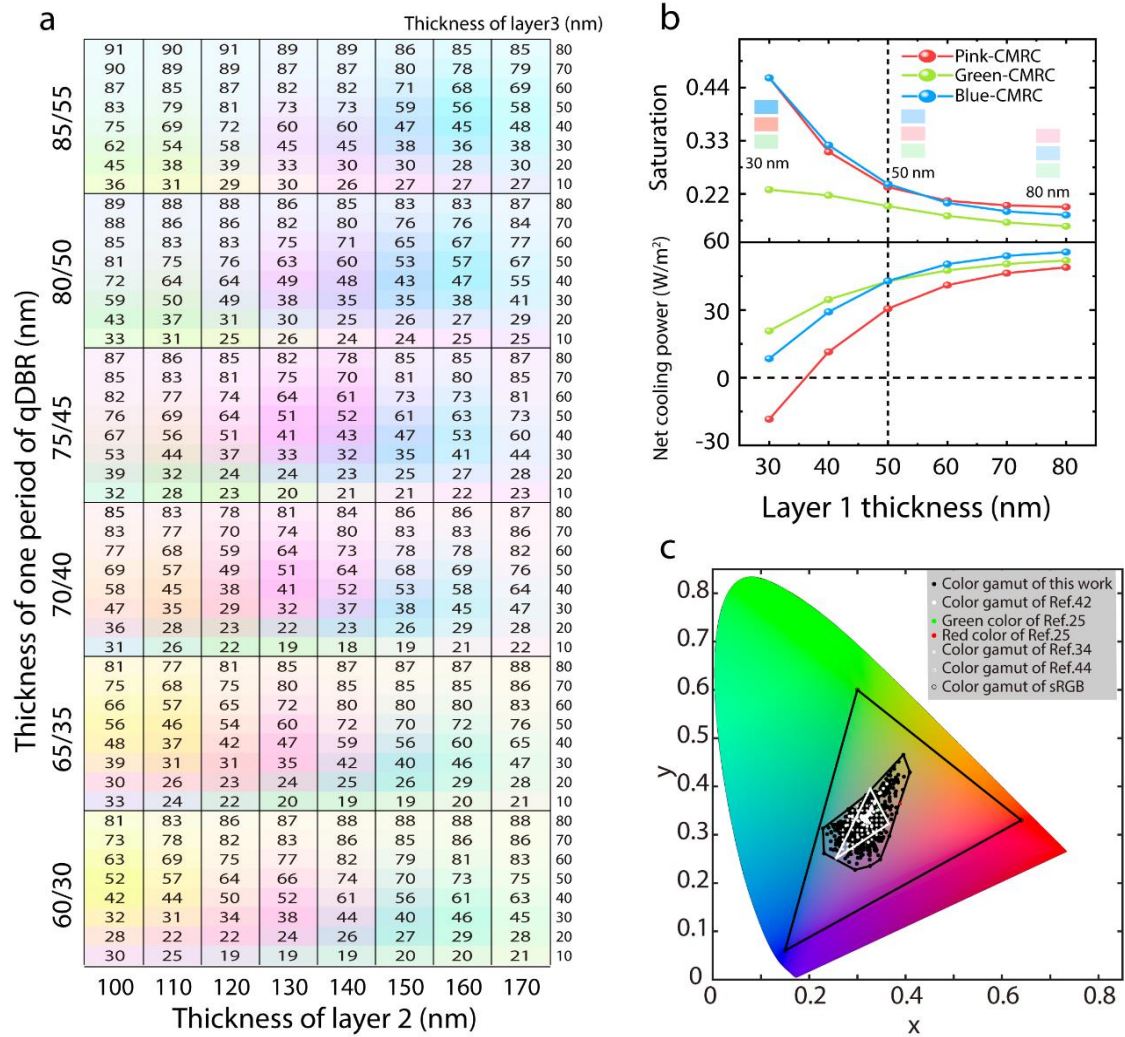

**Figure 3**. Colors and sub-ambient cooling ability of CMRCs. (a) Colors and net cooling power of CMRCs with a fixed Layer 1 thickness of 50 nm. (The thickness of the Layer 3 varies from 10 nm to 80 nm, whereas the thickness of Layer 2 varies from 100 nm to 170 nm at intervals of 10 nm, each of the MIM constitutes different qDBR-MIM photonic structures with six types of 2-periods qDBR structures. The thicknesses of each period layers are 60 nm/30 nm, 65 nm/35 nm, 70 nm/40 nm, 75 nm/45 nm, 80 nm/50 nm, 85 nm/55 nm, respectively.) (b) Color saturation and cooling performance of CMRCs as a function of the Layer 1 thickness and appearance of colors. (c) Color gamut of the CMRCs with positive net cooling power at ambient when the Layer 1 varying from 30 nm to 50 nm (the selection of other layer thicknesses are the same as Figure 3a), and the comparison between present work and previous works: the color gamut of the CMRCs in this work is the largest compared to all the colorful radiative

coolers with sub-ambient cooling ability in previous works including 1D-MIMIM structure,[44] the 1D-MIM structure,[42] the colorful coatings with core-shell nanoparticles,[34] and the perovskite NC-coated polymer.[25] The area of sRGB in the CIE plot is calculated to be 0.11205. the area of the color gamut of CMRCs with sub-ambient daytime cooling ability is 0.0198, which is 17.7% of the sRGB areal coverage, calculated by the method shown in Ref. (53)

**Experiments**

We further fabricated these CMRCs that show structural colors and sub-ambient cooling performance (Details of fabrication are provided in Supporting Information Section 6). We made three CMRCs with pink, green and blue colors as examples (Figure 4a). Note the colorless corners of samples are not broken but are caused by tape utilized to secure the sample during fabrication. Under cross-sectional scanning electron microscopy, we can observe all nine layers of a CMRC sample (bottom panel of Figure 4a), having consistent thicknesses with our design. Then, we compare theoretical and experimental spectra (Figure 4b-d). The experimental results agree well with theory in the wavelengths of 0.3~2.5 μm and 4~18 μm, which are mainly related to the radiative cooling performance. The absorption peak in the range of 2.5~4 μm likely results from the introduction of the hydroxyl group (–OH) during the fabrication. This range of wavelength has little effect on the solar radiation absorption due to the rare solar irradiance power in this wavelength range. Note the peak will only lead to an increase of 1.3%~1.5% (about $1.2\ W/m^2$) in the solar radiation absorption of CMRCs under $900\ W/m^2$ of incident solar intensity. Moreover, we perform a comparison between the radiative cooling performance of the fabricated pink-CMRC and existing colorful radiative coolers (Figure S11). Thanks to our novel design of coupled nanocavities, the pink-CMRC has the best cooling performance among existing colorful radiative coolers since it has the lowest solar radiation absorption and a considerable net thermal radiation in the wavelengths of the atmospheric window.

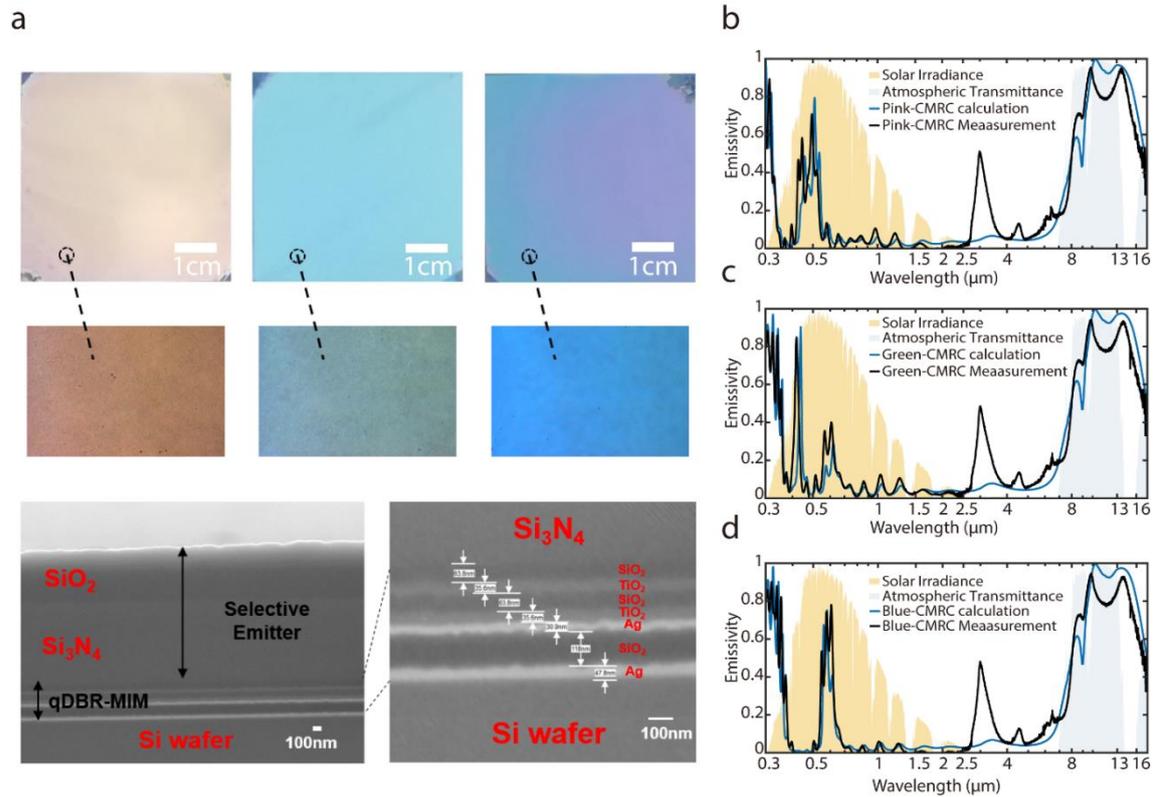

**Figure 4**. Characterizations of fabricated samples. (a) Top: Photographs of pink, green, and blue samples after fabrication. Center: Images of fabricated CMRC under an optical microscope. Bottom: Cross-sectional scanning electron microscopy image of Pink-CMRC. Each layer of CMRC can be clearly observed. The thickness of each layer in photonic structure is 63.8, 35.6, 63.8, 35.6, 30.9, 118, and 47.8 nm from the top to the bottom. (b, c, d) Comparison of measured and simulated emissivity of Pink-CMRC, Green-CMRC, and Blue-CMRC. The average solar reflection of three fabricated colorful samples is 89.65% (pink), 89.50% (blue), and 88.98% (green), respectively.

We experimentally measured the cooling performance of three CMRC samples on a flat roof in Shanghai, in mid-August 2021. The schematic of the custom radiative cooling measurement system is shown in Figure 5a, the experimental system consists of nine identical chambers utilized to place samples and shaded thermocouples for ambient temperature measurement (Details are provided in Supporting Information Section 8). We measured six samples during the outdoor experiment (Figure 5b), including three CMRCs a conventional daytime radiative cooler with broadband high reflectivity in the solar spectrum and white appearance, a broadband black emitter with high absorptivity within all wavelengths and black appearance, and a bare silicon wafer, which is the substrate of CMRCs and can be used to weigh the cooling performance of the CMRCs coating, were taken as comparative samples of CMRCs. The spectral reflectivity and emissivity of the conventional radiative cooler and broadband black emitter are given in Figure S12. The black emitter has broadband high emissivity within all wavelengths (0.3-18 μm) while the conventional radiative cooler has high reflectivity in solar spectrum and broadband high emissivity in MIR region.

As shown in Figure 5c and Figure 5d, during the hottest five hours of the day (from 10:00 to 15:00, 19th Aug), the blue, pink, and green CMRCs realized the average temperature drops of 4.3 °C, 4.3 °C, and 3.4 °C below the ambient temperature under an average solar irradiance of 724 W / m² and relative humidity of 47.11% (Figure S13). The maximum temperature drops of three samples are 6.5 °C, 6.4 °C and 5.4 °C, respectively. Meanwhile, the conventional radiative cooler achieved an average temperature drop of 9.0 °C and a maximum temperature drop of 11.8 °C during these five hours. In contrast, the bare Si wafer had an average temperature rise of 10.3°C and a maximum temperature rise of 16.0 °C, and the broadband black-emitter had an average temperature rise of 14.5 °C and a maximum temperature rise of 20.1 °C compared to the inner ambient temperature. Thus, three CMRC samples can present bright and vivid colors for aesthetics and succeed in the sub-ambient daytime radiative cooling under the sunlight in the hot summer of Shanghai. The nighttime radiative cooling performance of CMRCs is shown in Figure S14. Although the cooling performance is weakened due to the high relative humidity (~80%), our CMRCs can still realize sub-ambient cooling during the nighttime. The outdoor on-site measurements show the excellent cooling performance of the CMRCs and promising potential for radiative cooling applications.

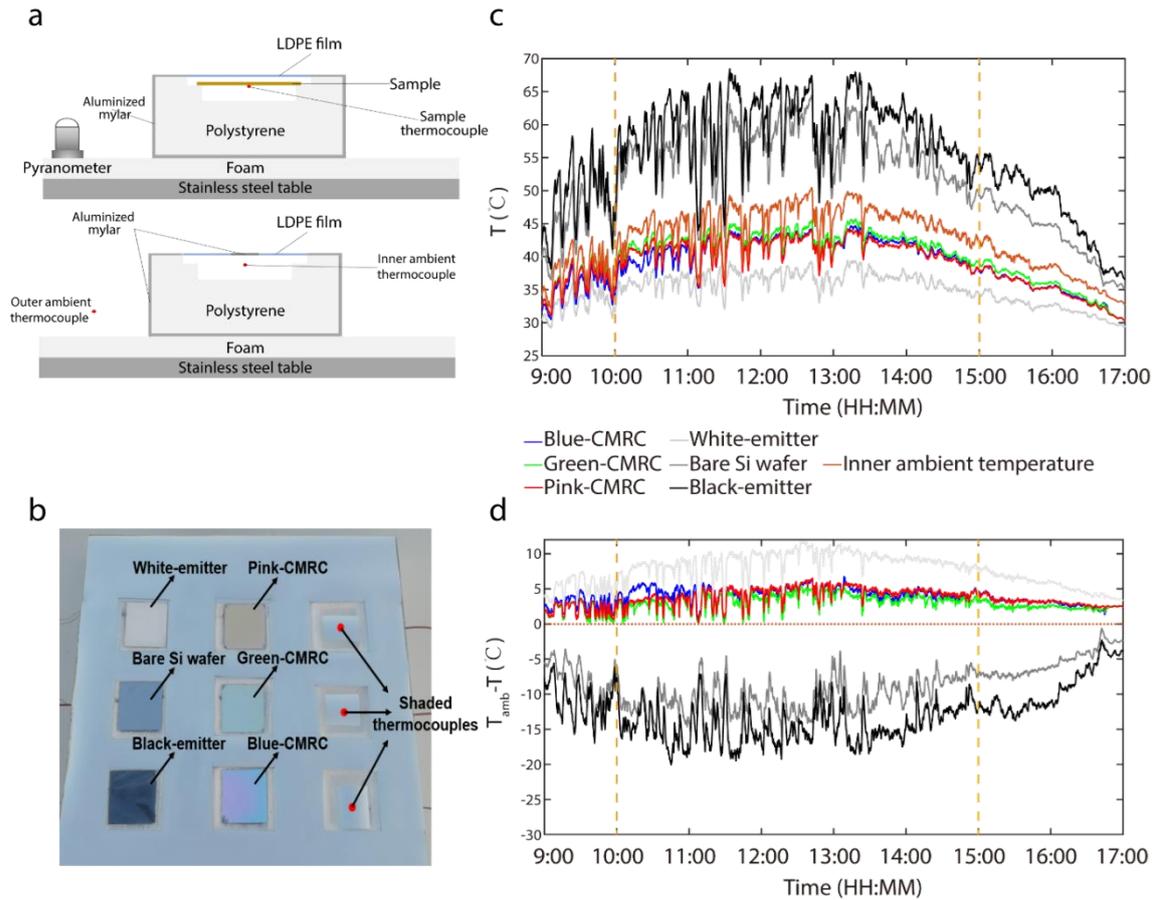

**Figure 5**. Passive daytime radiative cooling performance of fabricated CMRCs. This experiment was conducted from 9:00 to 21:00 on 19th Aug. (a) Schematic of the custom radiative cooling measurement system. (b) Photographs of the outdoor cooling temperature measurement system and six samples during measurement (19th, August, 2021, cloudless day). (c) Temperature measurements obtained on a rooftop from three CMRCs (red, green, and blue lines), a conventional white daytime radiative cooler (whitesmoke line), a broadband black-emitter (black line), and a bare Si wafer (grey line) compared with the inner ambient air temperature (orange line) during the daytime on 19th Aug. (d) Temperature difference between the inner ambient and six samples.

## Conclusion

We design a series of integrated structurally colorful CMRCs consisting of selective emitters and a photonic structure (qDBR-MIM). We demonstrate there are coupling effects between qDBR-Ag and MIM nanocavities. We can change geometric parameters to tailor the coupling effect and then create various structural colors with sub-ambient cooling performance. To realize desired spectrum for structural color, we can choose suitable original resonant frequencies of two independent structures at first by tuning the thicknesses of the qDBR structure and $SiO_2$ cavity in MIM, and then manipulate the coupling strength of two resonant modes inside the stacked photonic structure by tuning the thickness of central Ag layer. Based on these design rules, we can design CMRCs that show various colors with a larger color gamut than reported colorful radiative coolers. Meanwhile, the outdoor experiments show that our fabricated CMRCs have the largest temperature drops so far compared with the ambient temperature, which indicates the best sub-ambient cooling performance among all of reported colorful radiative coolers. The fabricated CMRCs can realize the average temperature drops of 3.4℃~4.4℃ and the maximum temperature drops up to 5.4℃~6.5℃ compared with the inner ambient temperature in the hottest 5 hours during daytime in the summer of Shanghai with an average solar irradiance of 724 W / $m^2$ and relative humidity of 47.11%. In addition, it is easy to produce our CMRCs through large-scale precision processing, like magnetron sputtering, vacuum evaporation and other fabrication processes. Our CMRCs can be utilized as energy-efficient colorful surfaces and have potentials in cooling of smartphone, wearable electronic/optoelectronic devices, and outdoor facilities.

## Supporting Information

## Competing interests

The authors declare no competing interests.

## Acknowledgments

We thank the financial support from the National Natural Science Foundation of China (Grants No. 51906144, No. 52120105009), Science and Technology Commission of Shanghai Municipality (No. 22ZR1432900, No.20JC1414800). The authors also thank the Center for Advanced Electronic Materials and Devices (AEMD) of Shanghai Jiao Tong University for the support in sample fabrications, and thank the Instrumental Analysis Center of Shanghai Jiao Tong University for the support in spectra measurement and scanning electron microscopy image.